# Assurance Cases as Foundation Stone for Auditing AI-enabled and Autonomous Systems: Workshop Results and Political Recommendations for Action from the ExamAI Project

Rasmus Adler [1][0000-0002-7482-7102] and Michael Klaes[1]

[1] Fraunhofer IESE, Kaiserslautern, Germany
`forename.surname@iese.fraunhofer.de`

**Abstract.**
The European Machinery Directive and related harmonized standards do consider that software is used to generate safety-relevant behavior of the machinery but do not consider all kinds of software. In particular, software based on machine learning (ML) are not considered for the realization of safety-relevant behavior. This limits the introduction of suitable safety concepts for autonomous mobile robots and other autonomous machinery, which commonly depend on ML-based functions. We investigated this issue and the way safety standards define safety measures to be implemented against software faults. Functional safety standards use Safety Integrity Levels (SILs) to define which safety measures shall be implemented. They provide rules for determining the SIL and rules for selecting safety measures depending on the SIL. In this paper, we argue that this approach can hardly be adopted with respect to ML and other kinds of Artificial Intelligence (AI). Instead of simple rules for determining an SIL and applying related measures against faults, we propose the use of assurance cases to argue that the individually selected and applied measures are sufficient in the given case. To get a first rating regarding the feasibility and usefulness of our proposal, we presented and discussed it in a workshop with experts from industry, German statutory accident insurance companies, work safety and standardization commissions, and representatives from various national, European, and international working groups dealing with safety and AI. In this paper, we summarize the proposal and the workshop discussion. Moreover, we check to which extent our proposal is in line with the European AI Act proposal and current safety standardization initiatives addressing AI and Autonomous Systems.

## 1 Introduction

Autonomous systems are systems that achieve predefined goals in accordance with the demands of the current situation without recourse to either human control or de-tailed programming [1]. The market prospects of these systems are promising. For instance, the global autonomous mobile robots market is projected to grow from $2 billion in 2021 to $8.7 billion in 2028 [2] and the global autonomous construction equipment market is expected to grow from $8.5 million in 2020 to $16.9 million in 2025 [3]. According to various roadmaps [1] [4], an autonomous system can be composed of several collaborating systems, like an intralogistics system that autonomously controls a fleet of various autonomous mobile robots and forklifts.

To put such systems on the European market, the Machinery Directive 2006/42/EC [5] has to be considered. The Machinery Directive is applicable for systems composed of several physical systems as it defines that "machinery" can be "assemblies of machinery". However, the Machinery Directive does not consider the term "autonomous" and thus not take into account that methods other than "detailed programming" are used to develop software. It demands only that a fault in the software of the control system should not lead to a hazardous situation. Safety standards that are harmonized with the Machinery Directive help practitioners to show that their systems are compliant with the Directive, giving them more detailed advice for specific aspects or types of machinery. They do consider software but focus on classical "detailed programming" and do not consider machine learning and other AI methods. Consequently, it is hardly possible to show compliance with a functional safety standard when AI is used to realize safety-relevant functionality. Autonomous systems with AI-based safety-relevant functions might thus not enter the European market even though they would increase safety and the efficiency of workflows.

Closing this gap is challenging for two reasons. First, it is challenging to develop suitable solutions for assuring the safety of autonomous systems and the correctness of their AI-based safety-relevant functions. Second, it is challenging to update the current normative and regulatory framework with respect to these solutions.

We propose the use of assurance cases to solve the first challenge, as these enable building up the required body of knowledge. We also propose integrating assurance cases into the current normative and regulatory framework. The two proposals are not novel, but we will derive them systematically from challenges with respect to the safety assurance of AI and autonomous systems. Moreover, we provide an overview of current research and standardization activities. One contribution is the formulation of the challenges that is intended to create awareness of pitfalls that may be encountered when enhancing the current normative framework based on the concept of Safety Integrity Levels (SIL), as currently applied in many safety standards. A second contribution is a discussion and evaluation of how assurance cases can help to avoid these pitfalls. The challenges and discussion presented in this paper are founded on a workshop where the problem statement was presented and assurance cases were proposed as a solution [6], as well as on a related whitepaper with policy recommendations [7].
The paper is structured as follows. First, we clarify core terms like "autonomous" or "AI" and provide some background on safety regulations and standards. Second, we state why the current safety standardization framework is in conflict with autonomy and AI. Third, we introduce assurance cases as solution. Fourth, we present to which extent this solution is being considered already in research, practice, standardization, and regulation. Fifth, we discuss and evaluate the solutions based on our workshop results. Finally, we present our conclusions and an outlook on future work.

## 2  Terminology and Background

### 2.1  Terminology on AI and Autonomous Systems

Because commonly accepted definitions are still missing – even though governments and companies are investing a lot in the topic of "AI" – it is quite unclear what actually belongs to the topic and whether the topic should not rather be called "autonomous systems". The German AI Standardization Roadmap [8] mentions the difficulties in defining AI and states that "in view of the difficulties in finding a generally accepted definition, this will not be done in this document". In the collection of definitions, the document presents five definitions of AI that are strongly related to the definition of an autonomous system in [1], which we presented in the beginning of the introduction.

To discuss challenges and solutions with respect to safety assurance, we need a term for systems that (1) behave in a very situation-specific way in a complex environment to achieve predefined goals, (2) do not or only partially depend on an operator to handle critical situations, and (3) are based on some software parts that can hardly be interpreted like software based on machine learning. In the following, we will use the term "autonomous system" for such systems, and we will use the term "AI" in order to refer to software parts that are not explicitly programmed and hard to interpret.

Our notion of AI as a special kind of software is in line with the AI definition of the European AI Act proposal. However, it emphasizes interpretability as the main aspect, thus we would not consider an interpretable expert system as AI, whereas the AI Act explicitly mentions expert systems in Annex 1, where it lists AI techniques and approaches.

Our notion of autonomous systems is in harmony with the definition we presented in the introduction. However, it supports the idea of having degrees of autonomy by being more or less dependent on an operator and by having more or less interpretability.

### 2.2  Background on Safety and Compliance

The Machinery Directive defines Essential Health and Safety Requirements (EHSR) that apply to all manufacturers who wish to put their products on the European market. Standards that are harmonized with the Machinery Directive support the demonstration of compliance with the ESHR. Of particular importance is EN ISO 12100, which provides a comprehensive list of different types of hazards and proposes the following three steps to address them.

The first step is called **inherently safe design.** In this step, hazards are avoided or related risks are inherently limited. For instance, inherently safe design could result in a design that avoids sharp edges in an engine with less power to inherently limit the speed of some machinery or some moving parts of it in order to limit the severity of a potential collision. In the second step, **technical protection mechanisms** are introduced. If the mechanisms are implemented by electric, electronic, or programmable electronic (software) means, a functional safety standard like IEC 61508 shall be applied to address potential malfunctions. For instance, a light barrier that triggers a shutdown is such a mechanism. In the third step, **user information about the intended use and the residual risk** is provided. For instance, warning signs are attached to the machinery in order to avoid critical deviations from the intended use.

These three steps shall be applied with an order of precedence. For instance, removing a hazard in step 1 has preference over addressing it with a protection mechanism in step 2. Similarly, a protection mechanism that detects and handles critical deviations from the intended use is better than a warning sign.

The application of the three steps results in a safety concept that needs to be enhanced with respect to critical malfunctions. A critical failure mode of any protection mechanism is obviously an omission because the protection mechanism cannot achieve the intended risk reduction if it omits acting as intended. However, other failure modes of a protection mechanism can also be safety-critical. For instance, a shutdown could be critical if it occurs in the wrong situation. Furthermore, malfunctions of functionalities other than protection mechanisms could be safety-critical.

The safety concept resulting from the 3-step method may prevent that these malfunctions lead to a hazardous situation, but this needs to checked. Thus, it is reasonable to assess every functionality with respect to safety. If it might be safety-critical, then every failure mode needs to be evaluated with respect to the risks that exist if the failure mode occurs. The higher the risks, the more safety measures should be applied to avoid the occurrence of the failure mode. In order to implement this relationship between risk and safety measures, functional safety standards define risk parameters, a mapping from possible parameter values to Safety Integrity Levels (SILs), and a mapping from SILs to safety measures to be applied (cf. Figure 1).

The latter mapping is intended to give safety engineers some freedom when choosing safety measures; that is, one SIL can be realized by applying different combinations of safety measures.

| | Technique/Measure | SIL 1 | SIL 2 | SIL 3 | SIL 4 |
|---|---|---|---|---|---|
| | Traditional Safety measure 1 | - | Recommended | Highly Recommended | Highly Recommended |
| | Traditional Safety measure 2 | Recommended | Recommended | Highly Recommended | Highly Recommended |

Software Engineering Steps (Requirements, Architecture,…)

*Figure 1 – Illustration of tables for SIL-dependent selection of safety measures*

## 3   Key challenges in Assuring AI and autonomous Systems 1

In this section, we will explain why the 3-step method and the SIL concept are generally not sufficient for autonomous systems and AI-based safety functions.

Inherently safe design is a reasonable approach for autonomous systems as long as it does not overly limit the utility of an autonomous system's situation-specific behavior. Unfortunately, this is very often the case. For instance, the inherent limitation of the speed and force of a gripper of a collaborative robot (cobot) reduces risks but lowers the performance of the cobot. Technical protection measures can act in a situation-specific manner in order to optimize the trade-off between safety and performance. For instance, a protection measure could limit the speed only if a human is close to the cobot. However, traditional protection measures are quite simple and not very situation-specific. Considering a mobile robot, a traditional measure would be a simple emergency stop rather than a situation-specific evasion maneuver. In many situations, situation-agnostic behavior typically results in unnecessarily low performance or availability. Situation-specific behavior can overcome this disadvantage. However, it is **challenging to assure safety if the safety concept comprises less inherently safe design, but some safety functions act in a very situation-specific manner in a complex context (problem 1)**. Situation-specific safety functions perceive the current situation and anticipate possible scenarios in order to perform in the best possible way from a safety perspective. The perception and anticipation capabilities go hand in hand with many implicit assumptions concerning the usage context, due to sensor limitations and calculation assumptions. For example, to conduct a successful evasion maneuver, the speed and direction of the human and other obstacles in the given context need to be anticipated, as must possible occlusions or other limitations in perception. It is **challenging to identify all these context assumptions and assure that they are fulfilled (problem 1.1)**. Another challenge stems from the open context. If the concrete operational context is not known or not constrained, safety engineers can hardly foresee and test all relevant operating conditions and scenarios. It is hard to assure that safety is achieved when the situation-specific safety functions act as specified, because the specified behavior might be unsafe in some **unforeseen edge cases (problem 1.2)**. For instance, autonomous mobile robots shall support various use cases and environments like open work yards with bicyclists and many other moving objects that need to be considered. If safety engineers are not aware of some types of moving objects, then the robots' evasive maneuvers might be inadequate for these kinds of objects.

Another issue is that **AI is not considered in functional safety standards and the related SIL-based selection of safety measures for software (problem 2)**. AI supports the implementation of situation-sensitive safety functions. For instance, data-driven models like deep neural networks for object classification are a powerful means for implementing the required perception capabilities. The development steps for such models differ significantly from the development steps for traditional software. In addition, the measures for fault avoidance, fault removal, fault tolerance, and fault forecasting also differ significantly. Consequently, the tables in safety standards like IEC 61508 that define which safety measures shall be applied in a certain development step depending on the SIL of the safety function are insufficient. As illustrated in Figure 2, a conceivable option could be to enhance the SIL concept by introducing development steps for a particular kind of AI and related tables that define which safety measure shall be applied depending on the SIL of the safety function.

| | Technique/Measure | SIL 1 | SIL 2 | SIL 3 | SIL 4 |
|---|---|---|---|---|---|
| Te... | AI Safety measure 1 | - | Recommended | Highly Recommended | Highly Recommended |
| 1 | AI Safety measure 2 | Recommended | Recommended | Highly Recommended | Highly Recommended |
| 1 | ... | Recommended | Recommended | Highly Recommended | Recommended |

AI-Engineering Steps (...,Data labeling,...)

*Figure 2 - Illustration of a possible SIL-dependent selection of safety measures for AI*

In doing so, the selection of new AI safety measures should be as effective for AI as the traditional ones for programmed software if the SIL is the same. This intuitive requirement is hard to fulfill for at least two reasons. First, there is **no explicit claim concerning the effectiveness of safety measures for software that has to be achieved (problem 2.1)**. Safety standards do not define metrics for measuring the effectiveness. There are probabilistic target values for safety functions depending on the SIL, but these values are only considered for dealing with random hardware faults, not for dealing with systematic faults, and all software faults are considered systematic faults. The second reason is that general **SIL-oriented claims concerning effectiveness can hardly be fulfilled for AI in practice (problem 2.2)**. For instance, the target failure probability of an SIL 4 safety function is $10^{-4}$ to $10^{-5}$ per demand. Data-driven models for common classification tasks generally have a likelihood of misclassification of between $10^{-1}$ and $10^{-3}$ (depending on the specific task), even if many state-of-the-practice measures have been applied to minimize the failure likelihood.

## 4 Advanced Assurance Cases for Autonomous Systems and AI

We propose assurance cases to deal with the presented challenges. In the following, we will present our proposal and discuss to which extent it solves or avoids these problems. First, we will introduce assurance cases as a general approach. Second, we will explain how assurance cases can support the systematic identification of context assumptions (problem 1.1). Third, we will present how assurance cases can be enhanced to deal with the problem of unknown edge cases (problem 1.2). Fourth, we will propose solutions to overcome the problem that there are no explicit claims concerning the effectiveness of safety measures (problem 2.2) and that SIL-oriented claims are hardly achievable (problem 2.2).

**Assurance Cases**

According to ISO/IEC/IEEE 15026, an assurance case is a reasoned, auditable artifact that supports the contention that its top-level claim (or set of claims) is satisfied, including systematic argumentation and its underlying evidence and explicit assumptions that support the claim(s). Figure 3 illustrates the general structure of an assurance case and the relationships of its key concepts. The left side shows that it basically comprises three elements: (1) some top-level claims saying that some objectives or constraints are fulfilled, (2) an argument or argumentation supporting the top-level claims, and (3) some evidence on which the argument is based.

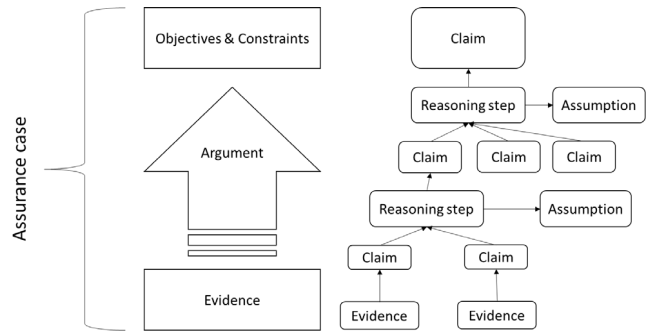

*Figure 3 – Illustration of an assurance case*

As shown in the right part of Figure 3, the argument can be structured by means of reasoning steps. In other words, the argumentation can be decomposed into different hierarchically structured arguments. However, we will use the terms "argument" and "reasoning step". Each reasoning step connects some sub-claims (or premises) with a higher-level claim (or conclusion). The reasoning step denotes that the higher-level claim is fulfilled if the sub-claims are fulfilled and if related assumptions hold.

In this paper, we focus on assurance cases where the top-level claim(s) relate to safety. In the case of a supplier of a safety component, the top-level claims refer to some safety properties of the component. In the case of a manufacturer of an autonomous system like a driverless transport system consisting of a fleet of autonomous mobile robots and other elements like cameras on the ceiling, the top-level claim refers to an acceptable level of safety for the intended use of the autonomous system. In the case of an operating company, the top-level claim refers to an acceptable level of safety for the concrete usage of the autonomous system.

Furthermore, we will focus on an approach that modularizes assurance cases and makes them composable. By means of this approach, a manufacturer can build their assurance case based on the assurance cases of their suppliers. The left part of Figure 4 illustrates such a modular assurance case. For instance, we consider a component that provides information on whether a worker is in a certain area. A top-level claim could be that the information is always correct if some assumptions hold. Some assumptions could refer to the absence of some failure modes of the input information. As illustrated in the right part of Figure 4, these assumptions create a link to the assurance case of the component that provides the required information. The manufacturer can ensure that the guarantee and demand interfaces of the different components fit to each other.

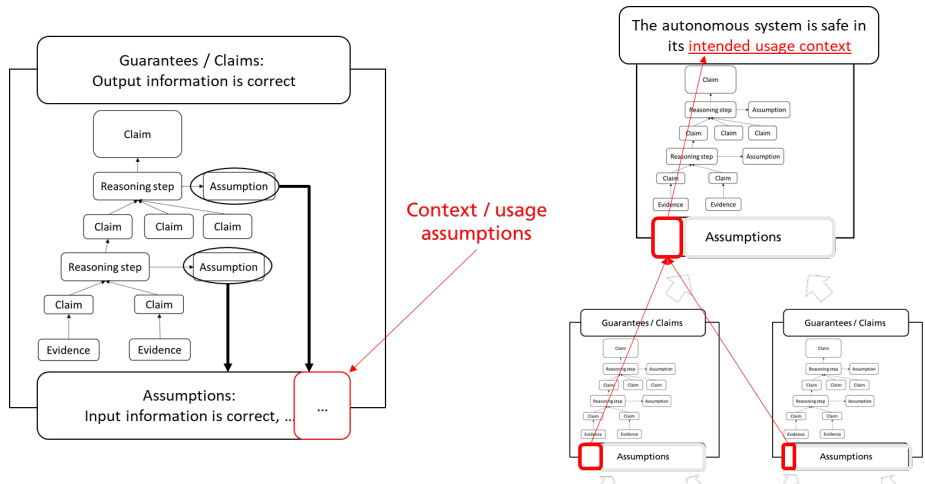

*Figure 4 – Illustration of the modularization of an assurance case*

The red circles in Figure 4 refer to assumptions that cannot be fulfilled by other components and constrain the usage. For instance, if a worker is detected by means of an infrared camera, then a fundamental assumption is that the worker has a certain temperature and this temperature is detectable by an infrared camera in spite of any clothes that the worker might wear or other circumstances. Accordingly, some restrictions concerning the clothes that the workers are allowed to wear could limit the intended usage.

A concrete solution that implements this modular approach is shown in Figure 5.

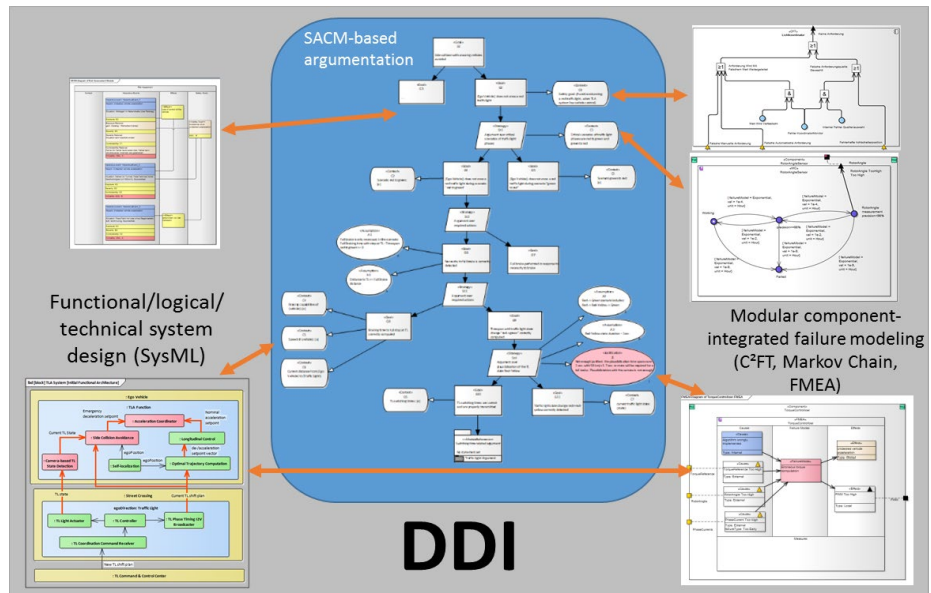

*Figure 5 Illustration of a Digital Dependability Identity (DDI)*

It illustrates a Digital Dependability Identity (DDI). The core of a DDI is a formalized assurances case. This core links to several models that are necessary to understand the safety argumentation; e.g., the hazard and risk assessment (HARA), the architecture including a functional, a logical, and a technical perspective, and different error-propagation models like component fault trees. By means of these models, a DDI describes in detail how assumptions have been derived, e.g., from a fault tree analysis. In addition, arguments for getting confidence in the completeness of the assumptions can be provided. Accordingly, these arguments are referred to as confidence arguments in [9].

**Identifying context assumptions with assurance cases**

Assurance cases are a means to explain how context assumptions have been derived and to provide arguments for their completeness. This enables assessing the argument and reasoning about its completeness. This kind of assessment differs from checking the requirements of a safety standard, which demands techniques for identifying context assumptions, because the argument that the implementation of some requirements leads to a complete set of assumptions is implicit. It is not based on the concrete results of the techniques demanded by the standard. An auditor can check all the work products of a safety standard, but the argument why these work products demonstrate the achievement of safety is implicit. This becomes a problem when fundamental assumptions that have been made when the standard was developed are no longer met; an example is the assumption that safety functions are rather simple and do not act in a very situation-specific manner in a complex environment. ISO 26262 uses the term safety-related function for a function that has the potential to contribute to the violation of a safety goal. The naming "safety-related" hints at functions that are not introduced for safety purposes like steering but that are safety-related. These functions have typically higher complexity than dedicated safety function that have been introduced in the second step of the 3-step method of ISO 12100.

However, functions comprising high situational awareness are not sufficiently addressed by the requirements of ISO 26262. Instead, the safety standard ISO/PAS 21448 (SOTIF – Safety of the Intended Functionality) has been developed in order to address performance limitations and insufficient situational awareness. The scope of ISO/PAS 21448 includes Advanced Driver Assistance Systems (ADAS) providing a low level of automation (i.e., SAE levels 1 and 2 according to SAE J3016), but the standard mentions that higher levels of automation might need additional measures. This means that the implicit argument would have some gaps. It would be hard to identify these gaps without making the implicit argument explicit. As illustrated in Figure 6, this can be done by means of an assurance case that uses the concrete outcomes from the requirements that have been implemented as evidences. We believe that such an assurance case can help to identify and address fundamental gaps of existing standards as well as gaps that are specific to the way the standards have been applied. In particular, it can help to access the completeness of the context assumptions by checking whether the manufacturer has considered all relevant aspects of the concrete usage context and whether the supplier has considered all relevant aspects of how their component is integrated into the autonomous system of the manufacturer.

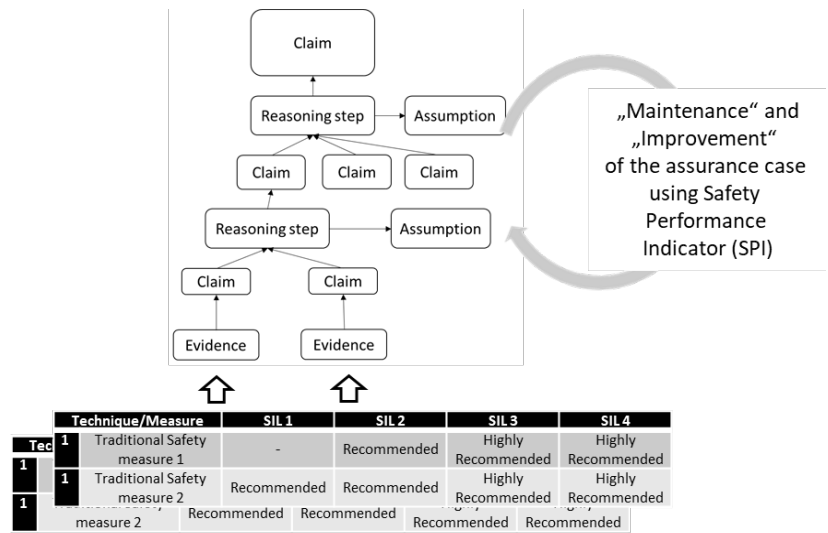

*Figure 6 – Relationship between an assurance case, safety standards, and SPIs*

**Identifying unknown edge cases with Safety Performance Indicators (SPIs)**

Figure 6 also illustrates the approach for attaching Safety Performance Indicators (SPIs) [10] to claims and assumptions in order to check after deployment of the system if they are true in application. For example, assumptions or claims concerning the frequency of situations could be measured by SPIs. This would contribute to the effectiveness of market surveillance. Furthermore, it would support the identification of unknown edge cases. An edge case that is relevant for a claim but that was not known and thus not considered in the underlying argumentation can be identified when measuring the fulfillment of a claim with SPIs. This approach is explained in more detail in [11] and considered in the ANSI/UL 4600 Standard for Evaluation of Autonomous Products [12]. The challenge of dealing with unknown edge cases or "unknown unknowns" is well known in the context of automated driving of road vehicles [13] because they operate in complex environments like cities. However, we see no reason why this approach should not be adopted for many kinds of machinery that are in the scope of the Machinery Directive [5] or the upcoming regulation on machinery products [14]. One example is autonomous mobile machinery that operates in open work yards where it is hard to foresee all possible harmful scenarios. The current draft of the regulation [15] already defines the Essential Health and Safety Requirement that "the risk assessment that manufacturers must carry out before the machinery is placed on the market / put into service will need to include also the risks appearing after the machinery is placed on the market due to its evolving and autonomous behavior". We see the use of assurance cases with SPIs as one powerful means for implementing this requirement.

**Claims and argument structure for AI-based safety functions**

Figure 6 does not show tables with SIL-dependent safety techniques/measures for AI as illustrated in Figure 2. Instead of introducing such tables, we propose developing a generic argument structure that can be instantiated or adapted. We provide various kinds of guidance encoded in this generic structure. First, we propose demonstrating that data-driven models are avoided as far as reasonably practicable. This means that it has to be argued that a data-driven model can hardly be replaced with traditional safety-critical software. Second, the consequences of failures of the data-driven model have to be reduced as far as reasonably practicable, e.g., by means of architectural measures such as fault tolerance mechanisms and plausibility checks. If these prerequisites are fulfilled, we propose arguing over the different elements of the autonomous system so that a claim for every data-driven model is established.

As illustrated in Figure 7, we would start with the claim that the data-driven model is sufficiently safe [37]. In the next step, we would make more concrete what we mean by "sufficiently" by means of two claims considering complementary risk acceptance criteria. The first claim is that the probabilities of the critical failure modes have been reduced as far as reasonably practicable. The second claim is that this reduction satisfies target failure probabilities. The first claim relates to safety measures during the lifecycle phases "Specification", "Construction", "Analysis", and "Operation", because in these phases, the failure probabilities can be reduced. The second claim relates to safety measures during the lifecycle phases "Specification", "Testing", and "Operation", because these phases are relevant to assuring that the residual failure probability is determined correctly. Because different phases rely on different datasets with different qualities, we also need to argue about the data used in these phases. Figure 7 illustrates only the general idea and not the argumentation with reasoning steps and concrete safety measures for data-driven models, as this is not relevant for discussing the role of assurance cases in regulation and standardization for AI systems and autonomous systems.

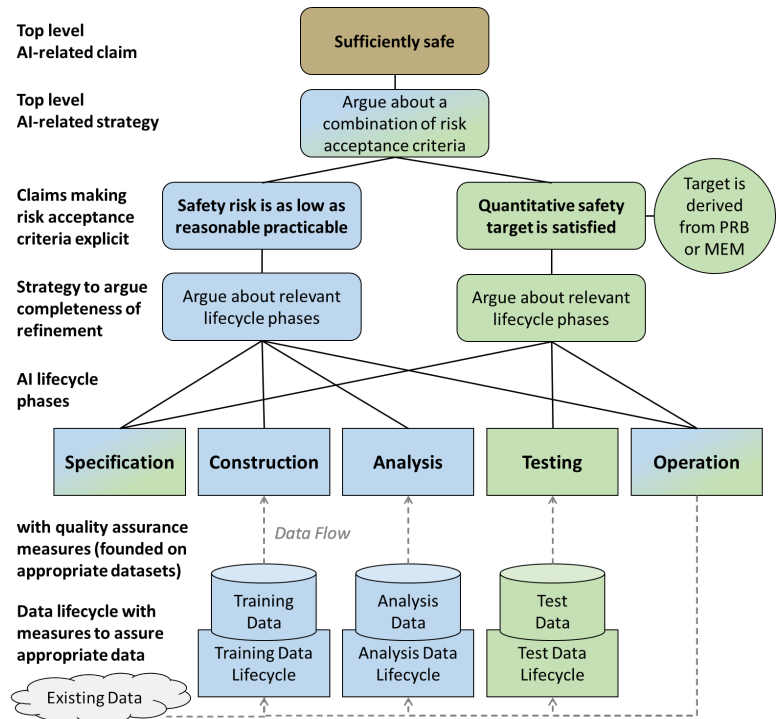

Figure 7 – Generic structure for arguing the safety of a data-driven model [37]

## 5 Assurance Cases in Research, Practice, Standardization, and Regulation

In the previous section, we sketched how the presented challenges concerning the safety assurance of AI and autonomous systems can be handled by means of assurance cases. We presented this sketch [16] in a workshop [17] in order to discuss it with relevant stakeholders and derive policy recommendations. Before summarizing the discussion and the recommendations in the next section, we will reflect on some activities and work in research, practice, standardization, and regulation that we considered when coming up with the proposal for focusing on assurance cases.

The solution sketch is related to German research projects focusing on automated driving. The Pegasus project [18] focused on the establishment of generally accepted quality criteria, tools, and methods as well as scenarios and situations for the release of highly automated driving functions. The overview picture of the Pegasus project illustrates that the complete testing and simulation framework is only for generating evidences for a safety argumentation [19]. The development of this safety argumentation was part of the project [20]. In the follow-up project V&V Methods [21], the work on this overall safety argumentation continues. Both projects did not consider in detail the safety argumentation for data-driven AI models. This challenge is considered in the project KI-Absicherung [22], which shall establish a stringent and provable safety argumentation with which AI-based functions can be secured and validated for highly automated driving. In all of these projects, the majority of German automotive OEMs and suppliers are part of the consortium. This indicates large consensus in the German automotive industry that assurance cases are a suitable means for assuring safety. This consensus is not limited to the German automotive industry. Several non-German standards for safety assurance cases are under development. The first one was UL 4600 [12]. The presented approach of attaching safety performance indicators to elements in an assurance case is already presented in UL 4600. ISO/NP PAS 8800 Road Vehicles - Safety and Artificial Intelligence [23] provides industry-specific guidance on the use of safety-related Artificial Intelligence/Machine Learning (AI/ML)-based functions in road vehicles. Its scope includes the derivation of evidence required to support an assurance argument for the overall safety of the system. This scope highlights the importance of arguing safety in an assurance case. Our projects with the automotive industry have shown us that assurance cases are already heavily used in practice. Standardization is thus a reasonable next step for documenting this acceptance of assurance cases in research and practice.

In aeronautics, assurance cases are also seen as a major solution approach for dealing with increasingly autonomous systems. The authors of the technical report "Considerations in Assuring Safety of Increasingly Autonomous Systems" from NASA believe that "an assurance case is the most suitable method to formalize the arguments and evidence … - providing a basis for assurance and certification of increasingly autonomous systems." The authors of another NASA report claim that assurances cases "address modern certification challenges" (cf. section 5.5) and explicitly mention "Managing Certification of Innovative Technology" (cf. section 5.5.4) as one challenge [24]. However, the European Union Aviation Safety Agency does not explicitly mention assurance cases in its guidance for machine learning applications [25] or its AI roadmap [26].

In the nuclear, railway, and medical industries, safety assurance cases are also well-known and established [24]. Furthermore, assurance cases are considered in defense, for example, in the DARPA Assured Autonomy Program [27] and in the assurance case framework from the Institute for Defense Analyses [28].

To tackle the challenges of autonomy, Asaadi et al. propose dynamic assurance using assurance cases that can be executed and evaluated during operation [29]. In the DEIS project [30], such machine-readable assurance cases called runtime DDIs are also considered. As we found several research activities in this and related fields, we initiated a workshop called Dynamic Risk managEment for AutonoMous Systems (DREAMS) in order to structure this new research field and create synergies [31]. As many of the solutions in this research field are, however, still in their infancy, we did not include them in our solution sketch.

Considering domains that are related to the Machinery Directive, like mining, agriculture, or manufacturing, assurance cases are considered in research but almost completely neglected in practice and standardization. The reason for this might be that (1) safety criticality is generally lower, (2) system complexity is lower, and (3) there is less need to use AI for realizing safety-critical functions compared to domains where assurance cases are already heavily used. The NASA/CR–2017-219582 report mentions these three challenges of modern certification that can be handled by means of assurance cases [24].

Autonomous systems that are in the scope of the Machinery Directive are addressed by the German application rule VDE-AR-E 2842-61 "Development and trustworthiness of autonomous/cognitive systems" [34]. This application rule has assurance cases as one of its key concepts but refers to them as trustworthiness assurance cases because the application rule considers not only safety but also other trustworthiness aspects.

To the best of our knowledge, assurance cases are not considered in regulation for AI and autonomous system. In particular, they are not considered in the current version of the European AI Act proposal [35]. We can only speculate why this is the case even though the goal-based nature of assurance cases is generally suitable for regulation, because regulation is more technology-neutral and more abstract than standardization. One reason could be that it would be too generic to demand only the development and auditing of assurance cases. However, an assurance framework like the Zenzic safety case framework [32] could introduce more concrete requirements concerning the development and auditing of assurance cases. Moreover, regulation could fix the first reasoning steps of the overall assurance case and define the resulting high-level claims as regulatory requirements that have to be shown. For instance, the AI Act follows a risk-based approach by defining three risk classes for AI systems and related measures for each class. An implicit argument or assumption here is that the measures are effective and do not lead to unnecessarily high effort that could be better used for other tasks to reduce risks. The authors of [33] put this into question; they propose more risk levels and a VCIO model that provides traces between Values, Critera, Indicators, and Observables. The VCIO model has similarities with an assurance case, as the traces between Observables and Values are similar to the traces between evidences and top-level claims or objectives.

## 6 Discussion about issues with assurance cases for certification

In context of the ExamAI project, we discussed the proposal of using assurance cases as a basis for the certification of autonomous machinery and AI-based safety functions with relevant stakeholders. We focused on machinery with smart manufacturing use cases due to the scope of the project. The discussion was, however, not limited to this focus and considered the horizontal regulation of AI with the European AI Act proposal. The results of the discussions in ExamAI have been published in a policy paper [7]. In the following, we will summarize the findings of the discussion in which we were involved.

A first finding concerns the issue of auditing an assurance case and coming up with objective criteria for assessing its acceptability. Ideally, different auditors would always arrive at the same results when assessing an assurance case. A process-based approach as presented in the Zenzic safety case framework [32] cannot solve this issue, as it defines only how the case is assessed but not under which conditions it is accepted.

Research proposes some techniques for quantifying confidence in assurance arguments but the work in [36] concludes that there is "no plausible justification for relying on one of these techniques in making decisions about which critical systems to deploy or continue to operate".

A related issue is that there are different guidelines for developing and structuring an assurance case. The proposal shown in Figure 7 and published in [37] does, for instance, differ from the AMLAS approach [38]. One goal of a current research project [39] is to achieve consensus on the relationship between the approaches and integrate them. It is currently being evaluated to which extent the approach shown in Figure 7 can be seen as an instantiation of the AMLAS approach, and to which extent the top-level claims in Figure 7 could be used as generic objective criteria for an assessment.

Another issue concerns the impact on industry and their concerns. One general concern is intellectual property (IP). As illustrated in Figure 4, our proposal implies that supplier, manufacturer, and operator work together on an integrated end-to-end assurance case. One could try to use modularization for IP protection purposes and hide the inner part of a modular assurance case. However, this approach has its limits. Different organizations have different knowledge and thus different capabilities for reviewing arguments and finding counterarguments. The conflict between IP protection and safety is a fundamental issue, but assurance cases suggest a solution that focuses on support for safety. In general, assurance cases are a very flexible and cost-efficient solution as one can really focus on what contributes to safety. However, a checklist-based approach requires less reasoning and can thus be faster and cheaper.

A related issue is the necessary level of expertise. Developing and auditing an assurance case for an AI-based safety function requires a high level of expertise with respect to safety assurance and data science. A skilled labor shortage exists in both areas and people that have a rich background in both areas are extremely rare. Thus, some guidance explaining what an assurance case for autonomous machinery should look like is necessary, not only for the purpose of objective assessment. Developing such guidance is a research task that has to consider acceptance by all relevant stakeholders, including industry, notified bodies, and public authorities like market surveillance authorities.

## 7   Policy recommendations and conclusions

In the ExamAI project, policy recommendations [7] have been derived to address the issues concerning certification based on assurance cases. In the following, we will summarize these three recommendations and draw conclusions.

The first recommendation is to create experimental spaces in which the solution approach can be applied and evaluated. The recommendation is in line with the "regulatory sandboxes" being considered in the European AI Act proposal in order to evaluate applications of AI systems. In the automotive domain, areas have been chosen where a specific infrastructure for automated driving is to be introduced, and many pilot projects are trying to introduce autonomous shuttle buses [40]. Such projects are essential to building the necessary body of knowledge for the safety assurance of autonomous systems. However, unlike in the automotive domain, there are only few projects in the smart manufacturing domain.

Assurance cases can be seen as a tool for building and communicating the body of knowledge. They help deal with empirical evidences but cannot generate them. On the one hand, autonomous machinery supports use cases that contribute to sustainability, well-being, and economic growth. On the other hand, building up the required body of knowledge for their safety assurance is very elaborative, expensive pioneer work. For this reason, means such as financial support have to make sure that all relevant stakeholders support this pioneer work even if they do not profit directly from the use cases.

A second recommendation concerns standardization support. AI-based safety functions and autonomous systems will be regulated and standards can significantly ease the task of triggering the presumption of conformity. However, it takes time and effort to achieve consensus on normative requirements for developing, monitoring, improving, and auditing assurance cases. Not all organizations can invest this time and effort. It is generally easier for large organizations than for small organizations. Furthermore, not all organizations that could contribute have sufficient incentives. For instance, some universities and research institutes with valuable expertise may struggle to find concrete benefits. Support for standardization should thus include means for accelerating the process and incentives for organizations to contribute with high expertise.

A third recommendation concerns a transculturation for many stakeholders. Safety culture is an established term that also hints at a certain mindset. As the AI research community and the safety research community have traditionally worked in isolation, their mindsets are different. However, both communities have to work together and find a consensus in order to develop solutions for the safety assurance of autonomous systems and AI-based safety functions. We see assurance cases as a suitable means for this collaboration because it demands explanatory power. It harmonizes with the high level of rigor in safety assurance but also with the empirical aspects of data science. The change in the safety community concerns the transition from checklist-based approaches to assurance cases. The change in the AI community concerns the introduction of a safety culture, related higher levels of rigor in engineering, and the representation of the level of rigor in an assurance case.

Please note that the policy paper [7] presents exactly these three recommendations but the descriptions differ. For instance, it considers industry, standardization organizations, and authorities of market surveillance when presenting transculturation with respect to assurance cases. The policy paper is in German and addresses politicians. The final report addresses a larger audience but is also in German. Our intention was to share the discussions within the ExamAI project and its conclusions with the growing Safety and AI research community.

We conclude that assurance cases are state of the art for assuring the safety of autonomous systems and AI-based functions, but this consensus in research is currently not reflected in relevant regulation and standardization. UL 4600, VDE-AR-E 2842-61, and ISO/NP PAS 8800 do consider assurance cases, but these standards are novel or under development and their acceptance by industry is currently unclear. Also, only ISO/NP PAS 8800 is at the international level, and its scope is limited to road vehicles.

The technical report ISO/IEC AWI TR 5469 "Artificial intelligence - Functional safety and AI systems", which is related to the international basic functional safety standard IEC 61508, is currently rather going in the direction of a rule-based approach.

## Acknowledgments

Parts of this work have been funded by the Observatory for Artificial Intelligence in Work and Society (KIO) of the Denkfabrik Digitale Arbeitsgesellschaft in the project "KI Testing & Auditing" (ExamAI) and by the project "LOPAAS" as part of the internal funding program "ICON" of the Fraunhofer Society. We would like to thank Sonnhild Namingha for the initial review of the paper.